The Binarity of Eta Carinae Revealed from Photoionization Modeling of the Spectral Variability of the Weigelt Blobs B and D [1]

Short title: Eta Carinae Binarity

E. Verner[2,3], F. Bruhweiler[2,3], and T. Gull[3]
kverner@fe2.gsfc.nasa.gov, fredb@iacs.gsfc.nasa.gov, and
theodore.r.gull@nasa.gov

---

Abstract

We focus on two Hubble Space Telescope/Space Telescope Imaging Spectrograph (HST/STIS) spectra of the Weigelt Blobs B&D, extending from 1640 to 10400Å; one recorded during the 1998 minimum (March 1998) and the other recorded in February 1999, early in the following broad maximum. The spatially-resolved spectra suggest two distinct ionization regions. One structure is the permanently low ionization cores of the Weigelt Blobs, B&D, located several hundred AU from the ionizing source. Their spectra are dominated by emission from H I, [N II], Fe II, [Fe II], Ni II, [Ni II], Cr II and Ti II. The second region, relatively diffuse in character and located between the ionizing source and the Weigelt Blobs, is more highly ionized with emission from [Fe III], [Fe IV], N III], [Ne III], [Ar III], [Si III], [S III] and He I.

Through photoionization modeling, we find that the radiation field from the more massive B-star companion supports the low ionization structure throughout the 5.54 year period. The radiation field of an evolved O-star is required to produce the higher ionization emission seen across the broad maximum. This emission region is identified with slow-moving condensations photoionized by the O star and located in the extended mass flow emanating from the B star primary. Comparison between the models and observations reveals that the high ionization region is physically distinct ($n_H \approx 10^7$ cm$^{-3}$ and $T_e \sim 10^4$K) from the BD Blobs ($n_H \approx 10^6$ cm$^{-3}$, $T_e \sim 7000$K).

Subject headings: atomic data – H II regions – ISM: individual (Eta Carinae Nebula) – line: formation – line: identification

## *1. Introduction*

Although Eta Carinae (hereafter Eta Car) has been extensively studied for more than 150 years, we know little about its central stellar source(s). Whether it is a single or multiple stellar system, it is hidden deep inside the dust-laden bipolar nebula, known as the Homunculus. Damineli (1996) found that the historical spectroscopic variability of Eta Car was consistent with a 5.54-year period, which led him to hypothesize that the central source was a stellar binary. Long-term observations showed the visual and near-IR spectrum to be a mixture of high and low ionization emission lines. The emission line spectrum changes from a state with high ionization lines to one where they are completely absent (Damineli et al. 1998). The variability in X-ray data can be explained by interacting stellar winds in a massive binary system (Corcoran et al. 1997; Pittard et al. 1998). The photoionizing flux from a B-star with an effective temperature of 15,000 K $\leq T_{eff} \leq$ 18,000 K cannot account for the high-ionization lines (e.g. [Ne III], [Fe IV], etc.) seen during the broad maximum away from the X-ray minimum, and implies that an additional, hotter photoionization source must also be present (cf. Verner et al. 2002).

Using HST/ Faint Object Spectrograph (FOS) observations, Davidson et al. (1995, 1997) suggested that the narrow and broad emission line components are formed in regions of different physical environments. The broad P-Cygni features were associated with the expanding wind of the central stellar source and scattered starlight within the Homunculus. On the other hand, the narrow emission lines were produced in the slow-moving nebular condensations (the Weigelt Blobs B, C, D and E, etc.) with a radial velocities $v_r \approx$ - 45 km s$^{-1}$.

Separation of the spectrum of the central source(s) from that of the Weigelt Blobs became possible with HST's near-diffraction limited performance and STIS's long, narrow aperture (Woodgate et al. 1997). High-spatial resolution, moderate spectral resolution spectra of central stellar source(s) of Eta Car and the Weigelt Blobs B and D (hereafter BD), located 0.1" and 0.2" respectively from the central stellar source (Gull et al. 2001), have been collected multiple times between early 1998 and 2004. As an initial step, we have used photoionization modeling to interpret STIS spectra obtained in March 1998, in a spectral minimum when low ionization nebular emission (Fe II, [Fe II], [S II], [N II]) dominated the spectrum (Verner et al. 2002). The February 1999 HST/STIS spectrum, centered on the central star and at the same position angle as in March 1998, exhibits, in addition to the low ionization lines seen in the March 1998 spectrum, lines of higher ionization species including [Ar III], [S III], [Si III], [Fe III], [Fe IV], and [Ne III] (e.g. Zethson 2001). In this paper we use our photo-ionization modeling of the minimum as a foundation and perform additional modeling of the spectroscopic maximum emission spectrum of February 1999.

The principal goal of this paper is to determine whether the presence of a hot secondary O-star in the stellar system is necessary to explain the variations seen in the emission spectrum of the BD Blobs near the spectroscopic minimum and maximum. Our approach to model the photoionization allows us to quantify the properties of the Eta Car central stellar source(s) through the observed properties of the BD Blobs. We select measurable, diagnostically important emission lines and use realistic ionizing fluxes of hot O stars deduced from model atmospheres (see Smith et al. 2002 and references cited therein) as inputs to the photoionization code CLOUDY[4] to obtain the best match with observations. As we will show, the high-ionization emission can be explained by photoionization from a hot luminous companion, while the cooler B star primary provides the radiative pumping via the stellar continuum flux to produce the strong Fe II and [Fe II] emission line spectrum.

---

[4] http://www.pa.uky.edu/~gary/cloudy

## 2. Observations of the BD Blobs

Both spectra of the BD Blobs were recorded with the HST/STIS, using the 52" X 0.1" aperture centered on Eta Car oriented at the identical position angle (P.A = 332.1$^o$). All the data were acquired using the CCD with the G230MB, G430M, and G750M gratings moved to standard grating settings yielding 0.05" pixel sampling and resolving power of ~ 6000 from 1640 to 10400Å. Each spectrum was recorded in a single HST visit. Significant savings in spacecraft resources were realized in 1999 by using continuous viewing zone orbits. Consequently, this cut the orbital overhead from five to two orbits. Figure 1 from Verner et al. (2002) shows the outline of the STIS aperture and its orientation relative to Eta Car (Component A) and the BD Blobs for these STIS exposures. The first spectrum was obtained in March 19, 1998 (STScI 7302, Davidson, PI) close to optical minimum, when Fe II and [Fe II] emission dominated the nebular emission spectrum. At minimum, lines of high component features such as [Ar III] and [Ne III] are absent. The second spectrum, obtained in February 21, 1999 (STScI 8036, Gull, PI) in the early part of spectroscopic maximum, clearly reveals additional nebular emission features. The details of the extracted spectra of the central stellar source and BD Blobs for the different visits, as well as a discussion of the extraction routines, can be found in Gull et al. (2001). Line identifications are presented in Zethson (2001).

Although many additional emission lines appear in the 1999 observations, not all of them can be reliably measured or used as physical diagnostics. Some emission features exhibit i.) unresolved blending, ii.) velocities associated with nebular structure that extends well beyond the BD Blobs. i.e. the Little Homunculus (Ishibashi et al., 2003), iii.) dust-scattered starlight (in the form of continuum and broad P-Cygni lines, iv) foreground nebular absorption at the other velocities, or v) superposition of emission originating from the resolved, extended stellar atmosphere (extended emission reaching 0.2 to 0.3" is apparent in the long aperture (Gull et al. 2001)). Consequently, we selectively use only well-isolated emission lines that are usable for physical diagnostics. These lines are listed in Table 1 with the following information: column (1) the observed ionic species, (2) the observed wavelength in the heliocentric frame, (3) the laboratory wavelength, (4) the deduced full-width half maximum of the feature (FWHM), (5) the observed intensity, (6) the reddening-corrected intensity, (7) the reddening-corrected intensity relative to [Ne III] λ3869, (8) the measured centroid line velocity, and (9) notes.

In our analysis of both emission regions we have used the flux ratio, [S II] λλ4069/ [S II] λ10322, to estimate extinction. Since these lines originate from a common upper level, their relative intensities can provide reasonable extinction estimates. We have adopted the [S II] transition probabilities of Czyzak & Krueger (1963) to obtain our reddening estimate. Using a Cardelli, Clayton & Mathis (1989) extinction curve and an $A_V/E(B-V) = 3.1$, we obtain an inferred $A_V = 0.467$ for the BD blobs. We have adopted this extinction for the region displaying the highly-ionized species as well.

We now focus on the most noticeable changes in the BD emission spectra between the minimum and the maximum, namely the change in ionization (section 2.1). Significant changes in the lower-ionization component, such as in the Fe II lines, are largely due to radiative pumping to upper levels, a process distinctly different from that which produces the features of the higher ionization species (Section 2.2).

### 2.1 Changes in Ionization Structure

The appearance of multiple high ionization emission lines in the spectrum of the Weigelt Blobs at the spectral maximum, in such species such as [Ne III], indicates an overall increase of the ionization in the observed emitting plasma. Figure 1 depicts the ionization range of the observed ions seen in both the 1998 and 1999 spectra (solid lines) and ions that appear only in the 1999 (maximum) spectrum (dashed lines). The ionization range for each ion is bounded on the low end by the ionization potential required to produce the ion and on the upper end by the ionization potential to the next ionic level. As Figure 1 illustrates, the 1998 (minimum) ionization ranges are systematically lower than those of 1999 (maximum). We call the lower ionization range group the "low ionization component" (LIC), and the higher ionization group the "high ionization component" (HIC), seen only during the broad maximum. The HIC includes elements seen throughout, namely Fe, Si, Ni, S, Cr, and N. The elements Al, C, Ar, and Ne are only seen during the maximum.

To demonstrate the change in strength of a HIC emission line in the BD Blobs, we plot the spectral region around [Ne III] $\lambda 3869$ for the two different epochs (Figure 2). Since the LIC is always present and does not vary significantly, we conclude that the HIC originates in a physically distinct region where the highly ionized species are due to ionization by a separate, second source only during the broad maximum.

### 2.2 Changes in Line Intensities

Most LIC lines between 1900–9500 Å show little change in flux between the minimum and the maximum. As an example, in Figure 3 we present the Fe II $\lambda 5020$ emission that results through pumping by strong continuum radiation from the B-star of the central source. An underlying broad component to the narrow nebular 5020 Å feature shows no change between 1998 and 1999. Although the LIC line of Fe II $\lambda 5020$ also shows little or no variability, the HIC lines of [Fe III] $\lambda 5012$ and He I $\lambda 5017$ are absent during 1998, but are present in the 1999 observations. These lines have no corresponding broad components. Finally, another important distinction between the LIC and HIC lines is that the FWHM of the HIC features, as presented in Table 1, are roughly half those of the LIC as given in Verner et al. (2002).

Very few lines disappeared in 1999, and those that did were very weak in 1998. Identified weak lines that disappeared belong to [Fe II] and [Cr II] and are likely due to relatively minor changes in ionization or excitation in the LIC emitting zone.

The profile shapes and strengths of the hydrogen Balmer and Paschen lines, arising from radiative recombination, are quite complicated and difficult to interpret due to multiple emission regions, dust-scattering of the complex, broadened stellar lines and intervening nebular absorptions. Thus, we do not attempt to model these lines, other than use them to provide upper limits for the model fluxes for the nebular emission.

Another group of LIC lines that show variability between 1998 and 1999 are associated with transitions that are pumped by H I Ly$\alpha$. They all respond strongly to changes in the Ly$\alpha$ flux. This group includes permitted lines of Fe II, Fe III, Ni II, and Cr II, especially noteworthy are the famous Fe II lines, $\lambda\lambda$2507 and 2509, demonstrated to be pumped by Ly$\alpha$ (Figure 4; Johansson & Letokhov 2003; and references therein). The intensities of Fe II $\lambda\lambda$2507, 2509 increased about 2.5-fold in the eleven month interval between the 1998 and 1999 observations. Besides these lines, the Fe II and the Cr II near-IR fluorescence lines (7895 − 10,111Å) are stronger by a factor of ~ 1.5 − 2 in 1999 (Zethson 2001). Since these lines are pumped via Ly$\alpha$, they do not necessarily behave as either HIC or LIC emission lines. Overall, the most significant changes in line intensities of the LIC between the minimum and maximum observations are in response to changes in stellar Ly$\alpha$ radiation that pump specific upper energy levels in Fe II.

### 2.3 Morphology of Emission Lines

Although there are many lines in the spectra of the BD Blobs, they can be separated by their various excitation conditions. The obvious differences in line shape depend upon the excitation conditions in the region of nebular emission. All continuum-pumped Fe II lines in the observations at near spectral minimum and maximum have both a narrow and broad component (For example, see Fe II λ 5020 in Figure 3). As mentioned above, the broad component is due to emission in the extended B-supergiant atmosphere of the central source. In contrast, the Lyα-pumped Fe II lines have no underlying broad component (see Figure 4 as an example). All high ionization lines that appear in 1999 have narrow profiles that are not associated with continuum-excitation of the B-star.

## *3. Physical conditions of high ionization structure*

### 3.1 Photoionization Models

The binary interpretation is based on observations in the visible, the radio and the X-Ray spectral regions (Damineli 1996, Damineli et al. 1998, Duncan et al. 1999, and Pittard et al. 1998). Non-binary interpretations include a periodic change in the atmosphere of a single star (Davidson et al. 1999) and a multiple star system requiring more than two stars (Lamers et al. 1998).

The most convincing evidence for stellar binarity comes from the X-ray variability, especially the reproducibility seen in the sharp, deep minimum occurring at 5.54 year intervals (Corcoran, 2004). The hydrodynamic modeling of Pittard & Corcoran (2002) gives a strong indication of the nature for the stellar secondary. Their model of colliding stellar winds leads to a synthetic spectrum that fits the X-ray emission obtained from *Chandra* spectra and qualitatively explains the observed X-ray flux variability. With a primary stellar mass-loss rate, $\dot{M}_1 \approx 10^{-3} M_\odot$

yr$^{-1}$ and a wind terminal velocity, $v_{\infty 1} \approx 5.5 \times 10^2$ km s$^{-1}$, the companion must have a stellar mass-loss rate, $\dot{M}_2 \approx 10^{-5} M_\odot$ yr$^{-1}$ and a wind terminal velocity, $v_{\infty 2} \approx 3 \times 10^3$ km s$^{-1}$. The derived values for the companion are consistent with an O supergiant or a Wolf-Rayet star. The deduced parameters of the orbit (Corcoran et al. 2001) imply that the periastron and apastron distances are roughly 1.5 A.U and almost 30 A.U, respectively (Pittard & Corcoran 2002). At closest approach, the secondary is well embedded, deep within the dense wind of the B star primary. The continuity equation, assuming isotropic flow, and constant nebulosity outflow ($v_{\infty 1} \approx 5.5 \times 10^2$ km s$^{-1}$) at larger distances, indicates that the ambient wind density around the secondary must be ~ $4.7 \times 10^{10}$ cm$^{-3}$ at 2 A.U. and near $2 \times 10^8$ cm$^{-3}$ at 30 A.U. Again, the continuity equation indicates that the characteristic density in the wind is proportional to r$^{-2}$, where r is the distance from the star.

However, there is evidence that the mass loss is not isotropic, but enhanced at the stellar poles. Moreover, inhomogeneties or condensations in the wind can greatly decrease the ambient wind density. Independent of the colliding wind modeling of Pittard & Corcoran (2002), we have adopted photoionization as the physical process responsible for the formation of the nebular lines. We now use the required UV-radiation to produce the HIC lines to also infer the nature of the stellar secondary in the Eta Car system. When we look at emission from the BD blobs, we see nebular emission produced at large distances from the primary, $10^{15} - 10^{16}$ cm (70 – 700 A.U.). Nonetheless, we have inferred the effective temperature and luminosity of the secondary star from photoionization modeling of the emission line spectra observed with HST/STIS.

In Verner et al. (2002), we analyzed the physical conditions during the spectroscopic minimum of the BD Blobs low-ionization structure, which is dominated by [Fe II] and Fe II emission. Detailed analysis of the emitting plasma suggests the BD blobs have a density in the range of $10^5 – 10^7$ cm$^{-3}$ and have physical dimensions of ~ $10^{15}$ cm. The bright BD Blob nebular emission at a distance of ~ $10^{16}$ cm from Eta Car requires a stellar effective temperature of $T_{eff}$ ~ 15,000K, with luminosity L ~ $10^{40}$ erg s$^{-1}$. During the broad maximum, a second, hotter stellar component is needed to explain the appearance of the high-ionization emission.

To model the emitting region, we must specify the luminosity, energy distribution in ionizing flux, the distance to the emitting region, and nebular abundances. The distance, in the plane of the sky, from Eta Car A to the BD Blobs, based upon the STIS imaging spectra, is estimated to be $3\times10^{15}$ cm $\leq R_{BD} \leq 3\times10^{16}$ cm. We have subsequently calculated a grid of predicted line fluxes varying properties of the secondary and using the photoionization code CLOUDY. The adopted spectral energy distribution for the secondary is given by WMBasic model atmospheres, which use line-blanketing and incorporate non-LTE and hydrodynamic flow (see Smith et al. 2002, and Lanz & Hubeny 2003).

In our modeling, we explicitly assume that the emitting region corresponds to the H II region surrounding the secondary. We assume the same range in total hydrogen density, $10^5 – 10^7$ cm$^{-3}$, as in our previous modeling. This preliminary approach assisted us in identifying the measurable line ratio, $\eta$ = [Ar III] $\lambda$7136/[Ne III] $\lambda$3869, as a gauge of the effective temperature of the secondary star. The threshold energies corresponding to the ionization potentials of Ar II-III and Ne II-III (Fig. 1) in the EUV and the shape of the stellar spectral energy distribution make $\eta$ very sensitive to the stellar flux distribution and the effective temperature of the hotter companion. The ratio, $\eta$, decreases with increasing $T_{eff}$ (e.g. $\eta \approx 1$ at $T_{eff} \approx 32,400$K and $\eta \approx 0.36$ at $T_{eff} \approx 37,200$K). Since the $\eta$ ratio depends on the reddening corrections, we recalculated the $\eta$ ratio assuming $A_v$=1.0 (also see Section 2). The $\eta$ value changes from 0.52 to ~ 0.33. The change in $\eta$ due to a change in $A_v$ from 0.467 to 1.0 is small, such that $T_{eff}$ is little affected (Table 3). The Ar/Ne abundance ratio also affects [Ar III] λ7136/[Ne III] λ3869. The photospheric abundances in massive B stars are expected to reflect those in the interstellar medium. However, the actual abundances in the ISM have been hampered by the uncertainties associated with how much is in the gas and dust phases of the interstellar gas. Recent work (Sofia & Meyer 2001) suggest that the solar composition should be a good proxy for the ISM, even though it appears to have general elemental enhancement of 0.2 dex relative to other stars in the solar neighborhood. Although η Car shows evidence of CNO-processing, this nucleosynthesis should not affect species heavier than N. Indeed, the general good agreement of the line intensities with predictions (Table 4) supports this.

Since the nebular features of the high ionization species result from photoionization by the hotter companion star, and since there are no significant changes in the LIC throughout the 5.5 year cycle, we initially ignore the LIC and focus on the HIC. If we assume all of the hydrogen ionization occurs within the Strömgren radius of the H II region surrounding the secondary, $r_{St}$, then using reasonable values of the number of hydrogen-ionizing photons, $Q(H^o)$, for different types of stars, we can estimate properties of the nebula using

$$r_{St}^3 n_H^2 = \frac{3}{4\pi}\frac{Q(H^o)}{\alpha_B}, where \quad \alpha_B = 2.6 \times 10^{-13}(cm^{-3}s^{-1}), n_H \text{ is the hydrogen density } (cm^{-3}).$$

Table 2 summarizes properties for each star used in our calculations: row (1) the identification of the model; row (2) the log of the number of total ionizing photons; row (3) $r_{St}^3 n_H^2$ values; row (4) effective temperature of the secondary star; row (5) log of mass-loss rate; and row (6) wind terminal velocity. The properties are collected from Smith et al. (2002), except for row (3), where $r_{St}^3 n_H^2$ values are calculated for different model atmosphere flux distributions representing O and Wolf-Rayet stars. The $Q(H^o)$ values are taken from Smith et al. (2002). The lower luminosity main sequence O stars (models 1 and 2) have the smallest Strömgren radius for a fixed hydrogen density. The largest Strömgren radius would be where the secondary is an O supergiant (models 3 and 4). With the presence of the dense wind of primary and the $r^{-2}$ falloff at larger distances from the primary, the Strömgren radius, $r_{St}$, *does* not represent the radius of a sphere, but a characteristic mean radius of the H II region of the secondary embedded in the dense, presumably neutral, wind.

The consequences of this modeling suggest that the HIC emission must arise in slowly-moving condensations embedded in the extended wind of the B star primary. If, instead, the emission arises in the expanding ambient wind and close to the primary, we would not see the narrow observed features (FWHM ~ 20–30 km s$^{-1}$) as indicated in Table 1 for the HIC features. The features would have presumably larger widths and projected radial velocities. For a density of $n_e$ ~$10^7$ cm$^{-3}$ in the vicinity of the O star secondary, one would obtain $r_{St}$ ~ 17 AU (Table 2). This dimension is smaller than the size of the orbit of the secondary. Also, as the secondary moves toward periastron, the O star and its H II region most likely moves away from the condensation giving rise to the HIC. Plus, the ambient density of the wind should increase dramatically, thereby decreasing the size of $r_{St}$. Thus, when the O star is close to the primary, the dense condensation cannot be photoionized by the O star. It is not clear if the HIC is associated with the bowshock produced by the stellar winds of the binary. However, the predicted intensities from the photoionization modeling are within a factor of two for most observed features (Table 4). For notable exceptions, see Section 3.3. This overall agreement argues favorably for photoionization by the O star producing the observed HIC emission.

For detailed calculations we have used the best available model of expanding non-LTE atmospheres. Although the new CMFGEN code of Hillier (private communication 2004) should yield better model atmosphere predictions, extensive grids for the stars of interest are not yet available. Since there is no direct information on photospheric abundances for the secondary, we used fluxes for stars with solar abundances. Calculations using the ionizing flux for a main sequence O star with $T_{eff}$ ~ 34,600K lead to an H II region with a dimensional mean radius that is ten times smaller than the derived distance to the BD blobs from the central stellar source. Thus, a main sequence O star with $T_{eff} \approx 34,600$K does not support an H II region that extends outward to the BD Blobs.

We have just used available flux distributions. The question is, "Is this for a WN or a WC –type Wolf-Rayet?" The biggest thing is how the C, N, and O opacities alter the EUV flux distribution. The main point is that one needs something luminous with a temperature in the range of 30,000-35000K. Whether it is a W-R or O star really doesn't matter. Meanwhile there is no real evidence of W-R star either in the optical or UV. The optical should show some very pronounced features. We hope to put constraints on this from more recent observations.

O supergiants have effective temperatures and sufficient luminosities to support the observed [Ar III] $\lambda 7136$ /[Ne III] $\lambda 3869 \approx 0.52$. A reasonable fit was achieved for an O star with $T_{eff} \approx 37,200$K and $\log (L_*/L_\odot) \approx 5.97$ assuming a hydrogen density of $10^7$ cm$^{-3}$. This O supergiant has a mass-loss rate $\log \dot{M} = -5.07$, with a stellar radius $R_* = 23.6\ R_\odot$ and a terminal velocity $v_\infty \approx 2 \times 10^3$ km s$^{-1}$. The [Ar III] $\lambda 7136$ /[Ne III] $\lambda 3869$ emission ratio for this model is $\eta \approx 0.4$. The model with temperature $T_{eff} \approx 34,600$K predicts $\eta \approx 1$. Thus, the effective temperature range $T_{eff} \approx 34,600 - 37,200$K for the secondary would explain the inferred [Ar III] $\lambda 7136$ / [Ne III] $\lambda 3869$ ratio in the 1999 observations.

The derived luminosity is large ($\log (L_*/L_\odot) = 5.97$), but significantly less than that of the primary. This value is sufficient to support the emitting region size of high ionization lines $\approx 10^{15}$ cm at hydrogen density $10^7$ cm$^{-3}$. Table 3 summarizes the predicted quantities: (1) Log of total luminosity in solar units, (2) Log of gravity, (3) Log hydrogen density, (4) column density, (5) minimum recombination time, (6) maximum recombination time (see section 3.2), (7) the size of the emitting region for given hydrogen density, (8) the average electron temperature, and finally, (9) $\eta$, the predicted [Ar III] 7136/ [Ne III] 3869 ratio. Models for main sequence or W-R stars can explain the value $\eta=1$, but their luminosities are too low to explain the size of the emitting region.

In Figure 5, we present the predicted ionization parameters for the regions supported by the primary star ($T_{eff}$ ~ 15,000K) and by the secondary star ($T_{eff} \approx$ 37,200K). The representative ionization fractions of a few ions (Ar III, Ne III, and Fe III) produced by photoionization by the secondary star are presented in Fig 5 (left). The extended Fe II structure (hereafter Fe II$^{(Pr)}$) produced by radiation field of the cooler primary is presented in Fig. 5 (right). The second component not only produces the HIC ions, but it may also produce Fe II (Hereafter Fe II$^{(S)}$ denotes the Fe II produced by the secondary, while Fe II$^{(P)}$ represents the Fe II due to the primary). Most of the Fe II$^{(S)}$ contributions are presumably lines that pumped via HI Ly$\alpha$.

A hot star that is much less luminous than the primary companion would support the HIC emitting region at a smaller distance from the binary system than that to the BD Blobs. At the same time, the HIC may be more extended than an ionized region supported by the cooler primary star. The model predicts that the variable emitting region of HIC must have a hydrogen density $n_H = 10^7$ cm$^{-3}$ with an electron temperature $T_e$ ~ $10^4$K (Table 3). Future high-spatial resolution observations that could separate the radiation originating in the HIC and the LIC would help to verify this hypothesis.

In our previous study (Verner et al. 2002), we found that the single 15,000K massive B-star could not provide enough nebular H I Lyα flux to explain the observed strong Lyα-pumped Fe II emission (e.g. λλ2507 & 2509). As a result, we postulated a very luminous and wide stellar Lyα to explain these Fe II strong lines. Our current model requires an O-supergiant as the hot secondary. The nebular Lyα intensity due to an O-supergiant is about hundred times larger than the continuum near Lyα. Whether this flux and wavelength span are sufficient to explain the observed fluxes of the Lyα –pumped Fe II lines, 2507 and 2509 Å is the subject of a future paper. If these Fe II lines are pumped by Lyα radiation from the secondary, they must vary throughout the 5.54-year period and should nearly disappear during the spectroscopic minimum. Future modeling of Lyα-pumped Fe II lines by the secondary star is needed to better constrain the Lyα-pumping contributions of the primary and secondary in Eta Car.

The total disappearance of HIC during the spectroscopic minimum indicates that the secondary, at closest perihelion, enters an ultraviolet absorbing gas that surrounds the binary system. The secondary must be completely embedded in the gas, which must attenuate or "squelch" the ionizing flux shortward of the Lyman-edge in all directions. If HIC lines in the emission spectra of Homunculus are due to photoionization from the hot stellar secondary, they should 1) disappear during the minimum event; and 2) fully recover when the second star reappears with sufficient time for reaching photoionization equilibrium. Pittard & Corcoran (2002) indicate that the X-ray flux is occulted during the minimum, but that the X-ray absorbing gas column increases ten-fold during the minimum. Whether this absorbing gas is due to the increased density of the wind near the secondary or whether additional contributions are needed is still unclear.

This binary system presents an unusual situation observationally. Normally, the presence of a luminous O star like that inferred in the Eta Car binary system would produce a large H II region. However, as mentioned earlier, this O star is largely embedded deep within the thick, mostly hydrogen-neutral stellar wind of a star undergoing mass loss at $\dot{M} = 10^{-3}$ M$_\odot$ yr$^{-1}$. Clearly, the H II region produced by the O star secondary is completely contained in the dense wind of the primary. Furthermore, given the expected r$^{-2}$ falloff in density expected at larger distances, the shape and dimension of the H II region will vary sharply with orbital distance, where the size of the H II region will become substantially smaller at smaller distances between the primary and secondary.

Perhaps the only unambiguous, direct spectral evidence seen in the STIS spectra of the O star is the presence of weak N V λλ1238, 1242 mass loss profiles. These wind features have long been known as UV spectral signatures of O stars (c.f. Heck 1987; Walborn et al. 1985). These features show absorption to at least –600 km s$^{-1}$. It is impossible to discern wind profile of N V at larger negative velocities due to the dramatically decreasing flux levels as a result of very strong H I Lyα absorption and an apparent drop-off in stellar continuum.

### 3.2 Ionic Recombination Times

We can estimate the ionic recombination time for the plasma in the Homunculus using temporal variations in spectra. X-ray emission spectra show a sharp disappearance of high ionization lines and their gradual recovery in a time scale of 2–3 months (Figure 1a from Ishibashi et al. 1999 and Figure 1 from Pittard & Corcoran 2002). Similarly, the ground-based spectra (Damineli, private communication, 2004) demonstrate the slow recovery of lines of high-ionization species (e.g. [Ne III] λ3869, [Ar III] λ7136).

Detailed calculations of recombination times require solving time-dependent equations for the ionization and line emission, which is beyond the scope of this paper. We have, nevertheless, provided a reasonable approximation for the hydrogen recombination time using our estimated densities. The recombination time is approximated by $\tau_r = 1/n_e \alpha_A$, where $n_e$ is the electron density and $\alpha_A$ is hydrogen recombination coefficient. The largest electron density for the one-zone model corresponds to the fully ionized plasma, $n_e \approx n_H$ Table 3 shows recombination times calculated for single zone model (column 6) and full model (column 7) for densities: $10^5$, $10^6$, and $10^7$ cm$^{-3}$ ($\alpha_A$ values are taken from Osterbrock 1974). The recombination time for the multi-zone O-supergiant model at $n_H=10^5$ cm$^{-3}$ is $\tau \sim 10^7$ seconds, which is about the same as the longest recombination time for the full zone O-supergiant model at $n_H=10^7$ cm$^{-3}$. However, these estimates have a large range of recombination times from $10^5$ s to $10^9$ s. Taking into account possible existence of regions with small column densities, or shock regions, the recombination times for high ionization lines are roughly four months at the maximum.

In addition to changes of high ionization structure, lines from singly ionized gas also demonstrate high variability. The time scales of their changes would be different due to their nature. Specifically, the Lyα-pumped Fe II lines, λλ2507 and 2509, where previous results (Verner et al. 2002) indicated strong Lyα flux, should change at a rate different from the recombination time. The strengths of these lines depend primarily on direct radiation, and should react to the secondary reappearance much faster. Therefore monitoring the fluxes of these lines before, during and after the minimum should provide much insight on the recovery of the UV flux from the secondary component. Many other lines, e.g. Fe III λ1914, are also pumped by Lyα and should provide parallel checks of the changes in the Lyα flux.

There are a number of discrepancies between our single-component B-star model and observations. Specifically some lines (see discussion on Fe II λλ 6433, 6457 (Verner et al. 2002)) would be better explained with a density ~ $10^7$ cm$^{-3}$ than with ~$10^6$ cm$^{-3}$. Ground-based observations by Damineli et al. (1998) demonstrated that these lines are variable while other Fe II lines (e.g. 5020Å) are not. Most lines are weaker during the minimum and stronger during the maximum. A possible explanation for such variations would be to assume that during the maximum most of emission comes from the HIC dominated by material with a density higher than that in the BD Blobs. The hydrogen density of $10^7$ cm$^{-3}$ would also give a recombination time of three months in Eta Car. Future detailed work is needed to study correlations of Fe II lines of different excitation conditions with ionization structure.

### 3.3 Constrains on Elemental Abundances

The chemical composition studies of the Eta Car (Davidson et al. 1986, Viotti et al. 1989, Dufour et al. 1997, and Hillier et al. 2001) showed that the abundances in the primary star and the surrounding nebulosities are high in nitrogen, and low in carbon and oxygen relative to the Sun. The BD Blob observations provide an independent approach to deriving abundances in Eta Car nebulosities and to understanding its current evolutionary state. Here, we concentrate on deriving abundances of the HIC seen in the 1999 observations. Since the neon abundance is largely unaffected in CNO-processing, unlike that of oxygen and carbon, we assume its abundance to be solar. Hence, we will use lines fluxes relative to [Ne III] λ3869 and thereby estimate abundances relative to Ne.

The observations and model fluxes are summarized in Table 4: column (1) gives the observed ionic species, column (2) the rest wavelength, (3) the observed reddening-corrected intensity relative to [Ne III] λ3869, (4) the predicted intensity relative to [Ne III] λ3869, and in

(5) ratio of theoretical to observed intensity. When we take into account the multiple uncertainties in setting the continuum, the complexity of the underlying spectrum shortward of 3500Å, the de-reddening correction for both intervening and internal dust, and line-blending, we consider a good fit to be when the model fluxes are within a factor of two of the measured fluxes.

### 3.3.1 Helium

Helium line fluxes change significantly during the 5.54-year period. During the 1998 observations only broad asymmetric features of He I lines (He I $\lambda 7066$ and He I $\lambda 7282$ are the strongest) are present in the BD Blob spectra. The broad profiles and extended nebular scales indicate they arise in the stellar wind or wind interaction of the stars rather than in the Weigelt Blobs. The broad components are also present in 1999, but narrow He I features are also present, coincident with the BD blobs. The strongest line that we were able to measure is at 7065 Å, but all He I lines accessible to STIS are weak. Yet, the He/H abundance ratio in the model had to be increased by 5-fold over solar to fit the observed line fluxes. Because there is large number of uncertainties (amount of dust, observational measurements, stellar continuum approximation for the secondary) we are conservative and tend to assume $5\times$ over solar. Formally $10\times$ fits better to the measured intensities.

No He II lines were identified in either the 1998 or 1999 epoch observations (Zethson 2001). This is in accord with our photoionization modeling, since there is a sharp dropoff in the radiation field shortward of the He II ionization edge in flux distributions of the O-stars and their model atmospheres.

### 3.3.2 Carbon, Nitrogen and Oxygen

Previous studies of the outer nebular structures have found very strong nitrogen lines, but virtually no carbon or oxygen lines (Zethson 2001).

Nitrogen abundance plays a fundamental role in Eta Car studies, because it is a key tracer of CNO processing. Nitrogen is overabundant in the star, Eta Car (Hillier et al. 2001) and the outer ejecta. The already strong [N II] lines, $\lambda\lambda 5756, 6585, 6549$ increased in strength from 1998 to 1999 to become among the strongest lines in the BD spectrum. A detailed examination of the STIS-observed [N II] emission at the position of the BD Blobs reveals a complex morphology; specifically [N II] $\lambda 6585$ exhibits a multi-peaked structure. The emission extends well beyond the positions of the BD Blobs indicating that line measurements are highly contaminated with emission from other nebular structures in the line-of-sight. We thus cannot use these line fluxes to derive accurate nitrogen abundance. Davidson et al. (1986) and Dufour et al. (1997) suggested that nitrogen is ~20 times solar abundance. In our models we used 10 times solar abundance for N as suggested by Hillier et al. (2001). In any event, there is a strong evidence for N overabundance in the ejecta of Eta Car.

The C III] intercombination line at 1908 Å is absent during in the 1998 data, but weakly present in the 1999 spectrum. Since C III] $\lambda 1906$ is not seen, the limit of the intensity ratio of $I(\lambda 1906)/I(\lambda 1908)$ gives an independent constraint on $n_e$, and implies an electron density $n_e \geq 10^4$ cm$^{-3}$ (Nussbaumer & Shield 1979). The HST/STIS spectra below 3500Å are dominated by very strong, complex absorption structures, with many velocity components that contribute to uncertainties in measuring this ratio. After C III] $\lambda 1908$, the strongest C lines predicted by O-star models are C II at $\lambda 1020$ and $\lambda 1335$. They and the C III at $\lambda 977$ are beyond the STIS CCD spectral range. Based upon our models fits to the 1999 observations, we conclude that carbon is $\sim 50 - 100$ times depleted in the BD Blobs.

Similar to the stellar spectrum, no emission line of any ion of oxygen is positively identified in the 1998 or 1999 BD Blob spectra (although we include possible lines in Table 2).

Even if these lines are present, they are very weak or badly blended with Fe II lines. Our model predicts oxygen in O I–III ionization stages. The strongest oxygen line predicted for the 1999 spectrum is [O II] λ2471. The model confirms that [O III] λ5007, if present, is severely contaminated by Fe II. We can only derive an upper limit to the oxygen abundance, being similar to that of carbon, is roughly a factor of 50 −100 below solar.

**3.3.3 Sulfur, Argon, Neon, Aluminum, Silicon and Iron**

Measurable lines of [S III], [Si III], [Ar III], [Ne III], [Fe III] and Al II] are present in the 1999 STIS spectrum of the BD Blobs. The fluxes of three [S III] lines (λ6313, λ9069, and λ9532) and four [Ar III] lines (λ3109, λ5192, λ7136 and λ7752) match well with our model calculations. Based on the model we expect that Ne and Ar are present in Ne I-III and Ar I-IV ionization stages during the maximum with the strongest lines originated from doubly-ionized state of each element. While the two strongest lines of [Ne III] ( λ3869 and λ3968) are present, there are no predicted Ne I and Ne II lines of detectable level in this spectral region. Similarly there are no strong Ar I, Ar II, or Ar IV lines expected in the studied region. The strongest two lines of singly-ionized form are [Ar II] at 6.989µm and [Ne II] at 12.81µm. The model predicts them to be weaker than the observed [Ar III] and [Ne III] lines and they are beyond the STIS spectral region.

The only identified emission of Si is Si III] at λ1891. Relative to the model flux, the measured line emission is weak by a factor of 10. Few silicon lines are predicted in the visible/ultraviolet spectral regions, so the silicon abundance is not well determined. If silicon is highly depleted in the gas, the most logical explanation is that silicate-based dust forms very close to the central source and locks up the silicon. Infrared observations (Smith et al. 2003) infer that silicates are a primary constituent of the emitting dust. The model predicts Si to be in Si II-IV ionization stages, where other silicon lines are at least 8 times weaker than the Si III] 1891 line. We used half of the iron solar abundance to obtain the fit between [Fe III] $\lambda 5270$ and $\lambda 4658$ with the observed values.

## *4. Summary*

Previously Eta Car has been studied assuming a wide range of scenarios. Rodgers & Searle (1967) favored a slow supernova description for this object. In contrast, Gratton (1963) had even suggested that Eta Car is a massive pre-main sequence star. Notwithstanding, most of the hypotheses supported the presence of an extremely massive star (Burbidge 1962; Burbidge & Stein 1970; Talbot 1971; Davidson 1971; Hoyle, Solomon & Woolf 1973; Humphreys & Davidson 1979; Davidson, Walborn & Gull 1982; Doom, De Greve & Loore 1986).

The suggestions that Eta Car was a stellar binary by Damineli (1996), Damineli, Conti & Lopes (1997) and Pittard et al. (1998) led us to look for the nebular properties produced by the presence of a hot companion. Our photoionization modeling of the spectrum of the Weigelt BD Blobs supports the stellar binary interpretation, and further implies that the primary and its companion both are very massive stars. The secondary is likely a massive, post-main sequence star with a lifetime of several times $10^6$ years (Andriesse, Packet & de Loore 1981) comparable to that suggested to the primary. Both stars should have the age of the open cluster Trumpler 16 of which Eta Car is presumed to be a member (Feinstein 1995).

Assuming that the secondary star is responsible for appearance and disappearance of the high ionization lines in the Homunculus Nebula, we have used realistic ionizing fluxes for an O supergiant and the photoionization code CLOUDY to obtain the best fit with observed emission line ratios and fluxes. Our conclusions are the following: 1) the LIC remains relatively unchanged; 2) The HIC is present during maximum and can be explained only by presence of a secondary, hot companion; 3) Lyα pumped lines increase in strength in the broad maximum by 2.5 times; 4) Fe II lines pumped by continuum show little change, and typically show broad underlying emission (presumably stellar); and 5) The Lyα pumped lines have no corresponding underlying broad component.

Among the measurable lines of high ionization species in the HST/STIS spectra we find that the [Ar III] λ7137/[Ne III] λ3869 ratio is very sensitive to the radiation field characterized by the stellar effective temperature. This ratio implies a stellar effective temperature, based upon our input model atmospheres, in the range 34,000K ≤ $T_{eff}$ ≤ 38,000K. The model atmosphere flux distribution for an O supergiant, $T_{eff}$ ≈ 37,200K and log ($L_*/L_\odot$) ≈ 5.97, realistically reproduces the ratio [Ar III] λ7137/[Ne III] λ3869 and explains the appearance of other high ionization species (e.g. C III], [S III], Si III], [Fe III] and He I) across the broad maximum. Such an O star can be represented by a spectral type and luminosity class of O7.5 I. The corresponding mass loss of the secondary star as inferred from the WM Basic model is within twenty percent of that independently derived by Pittard & Corcoran (2002) based on a wind collision scenario explaining the X-ray spectrum of Eta Car. The mass-loss rate of the secondary is about one hundred times smaller than that of the primary derived by Hillier et al. (2001).

Comparison of the periodically variable high ionization structure with the steady low ionization structure previously analyzed by Verner et al. (2002) suggests that they are probably distinct, but closely related, regions in the nebula. The high ionization structure is, likely embedded in the wind, located closer to the central source and it is identified with the H II region surrounding the secondary. This region is characterized by $T_e$ ~ $10^4$ K and log $n_H$ ≈ $10^7$ cm$^{-3}$. The low ionization structure is more distant, and cooler, $T_e$ ~ 7,000K, and log $n_H$ ≈ $10^6$ cm$^{-3}$. Evaluation of line ratios in the HIC suggests that carbon and oxygen are depleted 50 to 100 times relative to solar abundances. All measured helium lines can be fitted by models with the helium abundance increased by five times relative to solar. The model over predicts the strengths of [Fe III] and [Fe IV] lines by a factor of 2. Only one line of silicon has been observed, but it suggests a depletion factor of ten relative to solar. Line fluxes of other ions [S III], [Ne III], [Ar III] and Al II] are consistent with solar abundances.

The long-slit observations with HST/STIS at different angular positions relative to central source should be used to differentiate further the physical conditions of these nebulae. Detailed studies of velocity fields in the nebulosities would provide insight on the history of Eta Car. These observations are a close look at a unique massive star(s). These observations of Eta Car not only build a path to predict its future, but help us evaluate whether nucleosynthesis has a significant contribution from very massive stars (Heger & Woosley 2002).

The research of EV has been supported, in part, through NSF grant (NSF - 0206150) to CUA. Both, TRG and EV have been supported through the STIS GTO and HST GO programs. EV is grateful to R. Humphreys for helpful criticism and A. Damineli for useful discussions. We give special thanks to K. Ishibashi and N. Collins for data reduction and K. Feggans, D. Lindler, and T. Beck for their computational support.

Duncan, R. A., White, S. M., Reynolds, J. E. & Lim, J. 1999, Astron. Soc. Pacific Conf. Series 179, 54

Feinstein, A. 1995, RMxAC, 2, 57

Gull, T., Ishibashi, K., Davidson, K. & Collins N. 2001, in ASP Conf. Proc. 242, Eta Carinae and Other Mysterious Stars ed. T. Gull, S. Johansson, & K. Davidson (San Francisco: ASP), 391

Heck, A. 1987, in *Exploring the Universe with the IUE Satellite*, edited by Y. Kondo (Reidel, Dordrecht), p.121

Heger, A. & Woosley, S. E. 2002, ApJ, 567, 532

Hillier, D. J., Davidson, K., Ishibashi, K. & Gull, T. 2001, ApJ, 553, 837

Hoyle, F., Solomon, P. M. & Woolf, N. J. 1973, ApJ, 185, 89

Humphreys, R. & Davidson K. 1979, ApJ, 232, 409

Ishibashi K., Corcoran, M., F., Davidson, K., Swank, J. H., Petre, R., Drake, S.A., Damineli, A. & White, S. 1999, ApJ, 524, 948

Ishibashi K., et al. 2003, AJ, 125, 3222

Johansson, S. & Letokhov, V. 2003, A&A, 412, 771

Lamers H.G.J.L.M., Livio, M., Panagia, N., & Walborn, N. R. 1998, ApJ, 505, L131

Lanz, T. & Hubeny, I. 2003, ApJS, 146, 417

Nussbaumer, H., & Schild, H. 1979, A&Ap, 75, L17

Osterbrock, D. E. 1974, Astrophysics of Gaseous Nebulae

Pittard, J.M. & Corcoran, M.F. 2002, A&Ap, 383, 637

Pittard, J.M., Stevens, I. R., Corcoran, M. F. & Ishibashi, K. 1998, MNRAS, 299, L5

Rodgers, A. W. & Searle, L. 1967, MNRAS, 135, 99

Smith, L. J., Norris, R. P. F. & Crowther, P. A. 2002, MNRAS, 337, 1039
27

## 6. *Figure Captions*

Figure 1: Diagram of ionization potentials (eV) of ions observed at the same BD Blobs position in Eta Car close to minimum (March 1998) and during broad maximum (February 1999). The left side of diagram (solid lines) depicts the range of ionization potentials of all ions whose lines are always present in the spectra of the BD Blobs (LIC). The right side (dashed lines) presents the ionization potentials of ions whose lines that present during the broad maximum (HIC) but disappear during the minimum. All ions are ordered by increased lower limit of their ionization potentials in x-axis.

Figure 2: Extracted spectra from the two epochs (March 1998 and February 1999). The [Ne III] $\lambda3869$ is present during the broad maximum, which extends across five years of the 5.54 year period, but disappears during the several month-long minimum. Due to the high ionization potential (Ne II ~ 41eV < IP < Ne III ~ 63.45eV) the [Ne III] $\lambda3968$ line originates in nebular conditions that are different from that for Fe II lines.

Figure 3: Extracted spectra (March 1998 and February 1999) show one of the strongest Fe II lines $\lambda5020$. This line is pumped by UV continuum of the stellar primary ($T_{eff}$ ~ 15,000K). The line has a narrow component, which originates from the nebular Weigelt BD knots and a broad component, which arises from the extended stellar atmosphere. While the [Fe III] $\lambda5012$ and He I $\lambda5017$ appeared in 1999 and disappeared in 1998, the Fe II $\lambda5020$, both the nebular and stellar components, are unchanged.

Figure 4: On the left, variations of Fe II $\lambda\lambda2507$ and 2509 line intensities (1998 and 1999 observations) in the BD Blobs. For comparison, on the right, the same spectral region shows that the same lines in the RR Telescopii have different intensities.

Figure 5: On the left, ionization fractions of Ne III, Ar III and Fe III are predicted for $n_H$ = $10^7$ cm$^{-3}$ due to hot secondary companion ($T_* \approx 37{,}000$K). On the right, Fe II–III ionization fractions are calculated for $n_H$ = $10^6$ cm$^{-3}$ due to primary star ($T_* \approx 15{,}000$K), no Ar III or Ne III lines have been predicted.

## 7. Tables

Table 1. Selected Lines of Highly-Ionized Component Present in the BD Blobs (Feb. 1999).

| Spectrum | Rest Wavelength (Å) | Lab Wavelength (Å) | FWHM (km s$^{-1}$) | $I_{obs}$ 1.00E+12 (ergs cm$^{-2}$ s$^{-1}$) | $I_{corr}$ 1.00E+12 (ergs cm$^{-2}$ s$^{-1}$) | $I_{corr}$ /Ne III (3869) | δV (km s$^{-1}$) | Note |
|---|---|---|---|---|---|---|---|---|
| (1) | (2) | (3) | (4) | (5) | (6) | (7) | (8) | (9) |
| Si III] | 1891.650 | 1892.030 | 25.4 | 6.91 | 20.43 | 1.58 | -60.3 | |
| C III] | 1908.360 | 1908.730 | 33.6 | 0.75 | 2.22 | 0.17 | -58.2 | |
| [O II] | 2470.830 | 2471.090 | 17.0 | 0.28 | 0.73 | 0.06 | | 2470.970; blend with Fe II |
| Al II] | 2669.638 | 2669.948 | 26.0 | 1.54 | 3.84 | 0.30 | | |
| He I | 2829.589 | 2829.910 | 23.2 | 0.88 | 2.00 | 0.16 | -34.0 | |
| [Fe IV] | 2829.589 | 2830.190 | 23.2 | 0.88 | 2.00 | 0.16 | -63.7 | |
| [Ar III] | 3109.584 | 3110.080 | 30.0 | 0.22 | 0.47 | 0.04 | -47.8 | |
| [Ne III] | 3869.210 | 3869.850 | 32.9 | 7.35 | 12.91 | 1.00 | -49.6 | |
| He I | 3965.322 | 3965.850 | 19.6 | 0.50 | 0.86 | 0.07 | | |
| [Ne III] | 3967.903 | 3968.580 | 26.5 | 1.23 | 2.11 | 0.16 | -51.2 | |
| He I | 4026.674 | 4027.330 | 27.3 | 1.83 | 3.36 | 0.26 | -48.9 | |
| He I | 4144.532 | 4144.930 | 26.2 | 0.76 | 1.39 | 0.11 | | |
| He I | 4472.040 | 4472.730 | 24.8 | 4.42 | 7.72 | 0.60 | -46.3 | |
| [Fe III] | 4658.720 | 4659.350 | 24.5 | 5.23 | 8.88 | 0.69 | -40.6 | |
| He I | 4713.750 | 4714.470 | 20.8 | 0.93 | 1.57 | 0.12 | -45.8 | |
| He I | 4922.587 | 4923.300 | 24.7 | 1.22 | 2.01 | 0.16 | -43.5 | |
| [O III] | 5007.300 | 5008.240 | 19.2 | 0.85 | 1.38 | 0.11 | | blend with Fe II 5008.02 |
| He I | 5016.250 | 5017.080 | 25.7 | 4.29 | 6.98 | 0.54 | -49.6 | |
| He I | 5048.759 | 5049.150 | 19.6 | 0.67 | 1.09 | 0.08 | -23.2 | |
| [Ar III] | 5192.265 | 5193.260 | 16.9 | 0.14 | 0.22 | 0.02 | | |
| [Fe III] | 5271.153 | 5271.870 | 20.2 | 3.69 | 5.82 | 0.45 | -40.8 | |
| He I | 5876.280 | 5877.250 | 29.6 | 22.91 | 34.41 | 2.67 | -49.5 | |
| [S III] | 6312.900 | 6313.810 | 29.5 | 4.78 | 6.78 | 0.53 | -43.2 | |
| He I | 6678.936 | 6680.000 | 26.4 | 6.29 | 8.67 | 0.67 | -47.8 | |
| He I | 7066.100 | 7067.200 | 28.7 | 26.07 | 35.81 | 2.77 | -46.7 | |
| [Ar III] | 7136.664 | 7137.760 | 30.8 | 4.89 | 6.69 | 0.52 | -46.1 | |
| He I | 7282.178 | 7283.360 | 29.6 | 2.88 | 3.91 | 0.30 | -48.7 | |
| [O II] | 7321.648 | 7322.000 | 21.1 | 0.49 | 0.66 | 0.05 | | Blended with 7320.99 N I |
| [Ar III] | 7752.079 | 7753.240 | 33.7 | 1.40 | 1.85 | 0.14 | -44.9 | |
| [S III] | 9069.724 | 9071.110 | 22.0 | 2.18 | 2.67 | 0.21 | -45.8 | |
| He I | 9464.524 | 9466.190 | 21.2 | 0.99 | 1.18 | 0.09 | -52.8 | |
| [S III] | 9532.120 | 9533.234 | 27.3 | 7.71 | 9.18 | 0.71 | -35.1 | |

Table 2. Predicted $r_{St}^3 n_H^2$ Values of H II Region near Eta Car due to Secondary Star.

| Model # | | 1 | 2 | 3 | 4 | 5 | 6 |
|---|---|---|---|---|---|---|---|
| ID* | (1) | OB #5 | OB #6 | OB #16 | OB #17 | WN #3 | WN #4 |
| Log Q(H°)* (photon s$^{-1}$) | | 48.7 | 48.5 | 49.6 | 49.5 | 49.1 | 49.2 |
| $r_{St}^3 n_H^2$ (n in cm$^{-3}$, $r_{St}$ in pc) | (2) | 60.66 | 60.46 | 61.56 | 61.46 | 61.06 | 61.16 |
| T$_*$ (kK) | (3) | 37.2 | 34.6 | 37.2 | 34.6 | 35.0 | 40.0 |
| Log $\dot{M}$ (M$_\odot$ yr$^{-1}$) | (4) | −6.47 | −6.64 | −5.07 | −5.19 | −4.86 | −4.83 |
| v$_\infty$ (km s$^{-1}$) | (5) | 2100 | 1950 | 1980 | 1950 | 1240 | 1370 |

* Model numbers (IDs) and Log Q(H°) are taken from Smith et al. 2002 (Tables 1-3). For comparison, the estimated total ionizing photons, for the primary B star is Log Q(H°) = 46.15.

Table 3. Physical Conditions in the BD Blobs (Feb 1999)

| Model | Log $(L_*/L_\odot)$ [a] | Log(g) [a] | Log $n_H$, (cm$^{-3}$) | $N_e$ x $10^{22}$ (cm$^{-2}$) | $\tau_{recmin}$ ($10^7$ s) | $\tau_{recmax}$ ($10^7$ s) | R ($10^{15}$ cm) | $T_e$ ($10^4$ K) | [Ar III] 7136/ [Ne III] 3869 |
|---|---|---|---|---|---|---|---|---|---|
| (1) | (2) | | (3) | (4) | (5) | (6) | (7) | (8) | (9) |
| O supergiant | 5.97 | 3.4 | 5.0 | 1.36 | 1.34 | 21.3 | 136 | 1.09 | 0.62 |
| | | | 6.0 | 2.28 | 0.13 | 3.97 | 22.7 | 1.21 | 0.51 |
| | | | 7.0 | 1.39 | 0.01 | 1.09 | 1.38 | 1.17 | 0.36 |
| main sequence star | 5.3 | 4.0 | 5.0 | 0.64 | 1.34 | 20.9 | 64.4 | 1.00 | 1.6 |
| | | | 6.0 | 0.77 | 0.13 | 4.4 | 7.73 | 1.3 | 1.23 |
| | | | 7.0 | 0.16 | 0.01 | 1.09 | 0.16 | 1.13 | 1.27 |
| W-R star | 5.64 | 3.7 | 5.0 | 0.96 | 1.34 | 21.2 | 96.4 | 1.05 | 0.8 |
| | | | 6.0 | 1.43 | 0.13 | 4.45 | 14.3 | 1.18 | 0.63 |
| | | | 7.0 | 0.58 | 0.01 | 1.12 | 0.58 | 1.12 | 0.48 |

[a] To explain the appearance of high ionization lines we have performed the calculations (O and C are depleted 100 times compare to solar abundances) for different hydrogen densities using flux of O-type star from WMBasic grid (Smith et al. 2002). All stars has log $(T_{eff}) \approx 4.57$. The total size of emitting region predicted in calculations that stopped at temperature $\approx$ 4,000K.

Table 4 Comparisons between Observations and Model Predictions [a]

| Spectrum | Rest Wavelength (Å) | $I_{cor}$/NeIII(3869) | $I_{th}$/NeIII(3869) | $I_{th}/I_{obs}$ |
|---|---|---|---|---|
| (1) | (2) | (3) | (4) | (5) |
| Al II] | 2669.638 | 0.298 | 0.288 | 0.97 |
| [Ar III] | 3109.584 | 0.036 | 0.045 | 1.26 |
| [Ar III] | 5192.265 | 0.017 | 0.018 | 1.05 |
| [Ar III] | 7136.664 | 0.518 | 0.232 | 0.45 |
| [Ar III] | 7752.079 | 0.143 | 0.056 | 0.39 |
| C III] | 1908.360 | 0.172 | 0.104 | 0.61 |
| [Fe III] | 4658.720 | 0.688 | 0.888 | 1.29 |
| [Fe III] | 5271.153 | 0.451 | 0.506 | 1.12 |
| [Fe IV] | 2829.589 | 0.155 | 0.243 | 1.56 |
| He I | 2829.589 | 0.155 | 0.012 | 0.07 |
| He I | 3965.322 | 0.066 | 0.075 | 1.13 |
| He I | 4026.674 | 0.26 | 0.198 | 0.76 |
| He I | 4144.532 | 0.107 | 0.033 | 0.31 |
| He I | 4472.040 | 0.598 | 0.345 | 0.58 |
| He I | 4713.750 | 0.122 | 0.071 | 0.59 |
| He I | 4922.587 | 0.155 | 0.092 | 0.60 |
| He I | 5016.250 | 0.54 | 0.193 | 0.36 |
| He I | 5048.759 | 0.084 | 0.016 | 0.20 |
| He I | 5876.280 | 2.665 | 0.903 | 0.34 |
| He I | 6678.936 | 0.672 | 0.215 | 0.32 |
| He I | 7066.100 | 2.774 | 0.885 | 0.32 |
| He I | 7282.178 | 0.303 | 0.079 | 0.26 |
| He I | 9464.524 | 0.092 | 0.017 | 0.19 |
| [Ne III] | 3869.210 | 1 | 1 | 1.00 |
| [Ne III] | 3967.903 | 0.163 | 0.3 | 1.84 |
| [O II] | 2470.830 | 0.056 | 0.064 | 1.13 |
| [O II] | 7321.648 | 0.051 | 0.042 | 0.83 |
| [O III] | 5007.300 | 0.107 | 0.021 | 0.19 |
| [S III] | 6312.900 | 0.442 | 0.816 | 1.85 |
| [S III] | 9069.724 | 0.207 | 0.424 | 2.06 |
| [S III] | 9532.120 | 0.711 | 1.061 | 1.49 |
| Si III] | 1891.650 | 1.582 | 1.183 | 0.75 |

[a] The Log gas-phase elemental abundances are as following:
H : He : C : N : O : Ne : Na : Mg : Al : Si : S: Cl : Ar : Ca : Fe = 0.00 : - 0.30 : - 5.45 : - 3.03 : - 5.13 : - 3.93 : - 5.69 : - 4.42 : - 5.53 : - 5.45 : - 4.79 : - 6.73 : -5.40 : - 5.64 : - 4.79

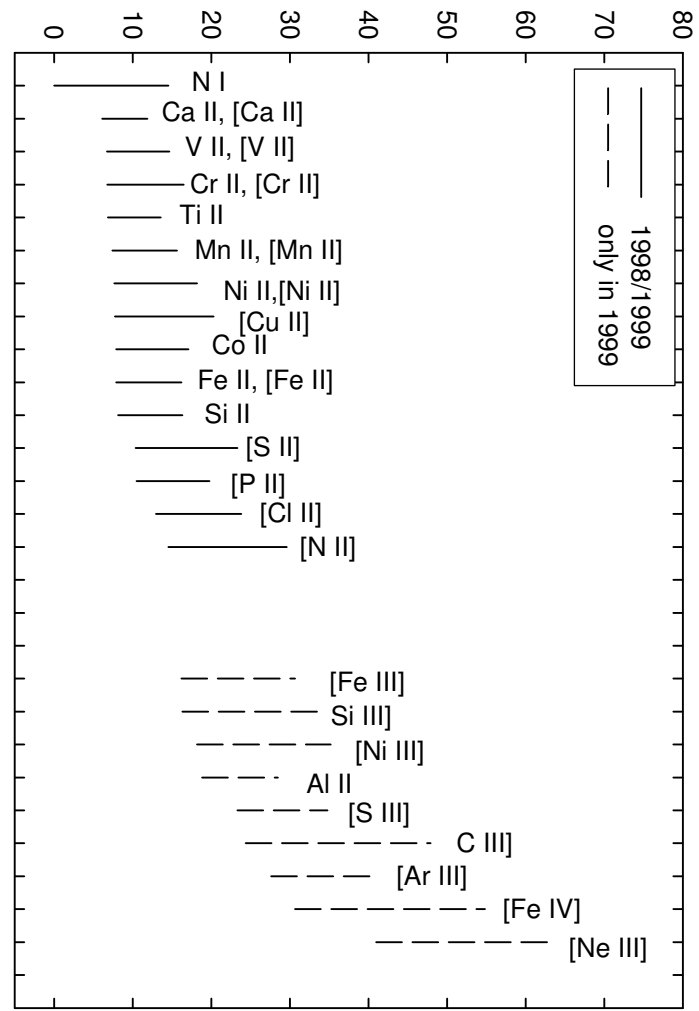

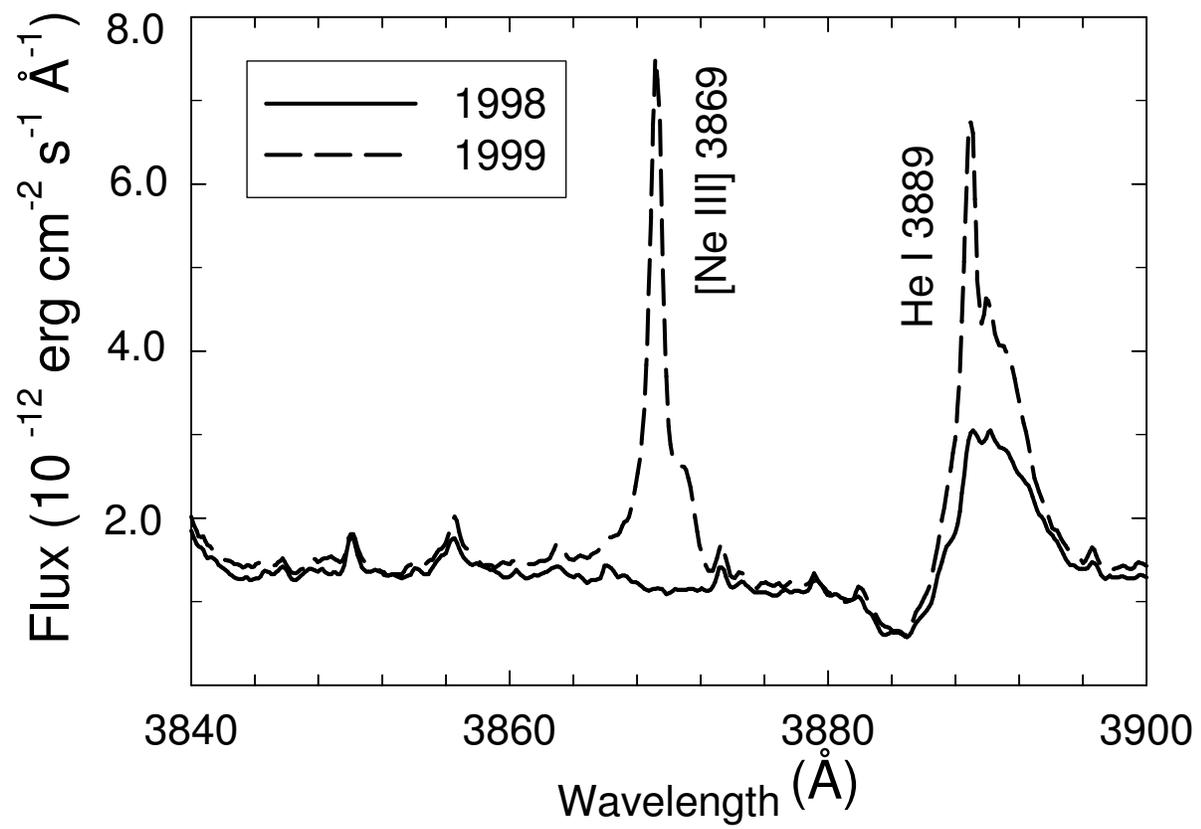

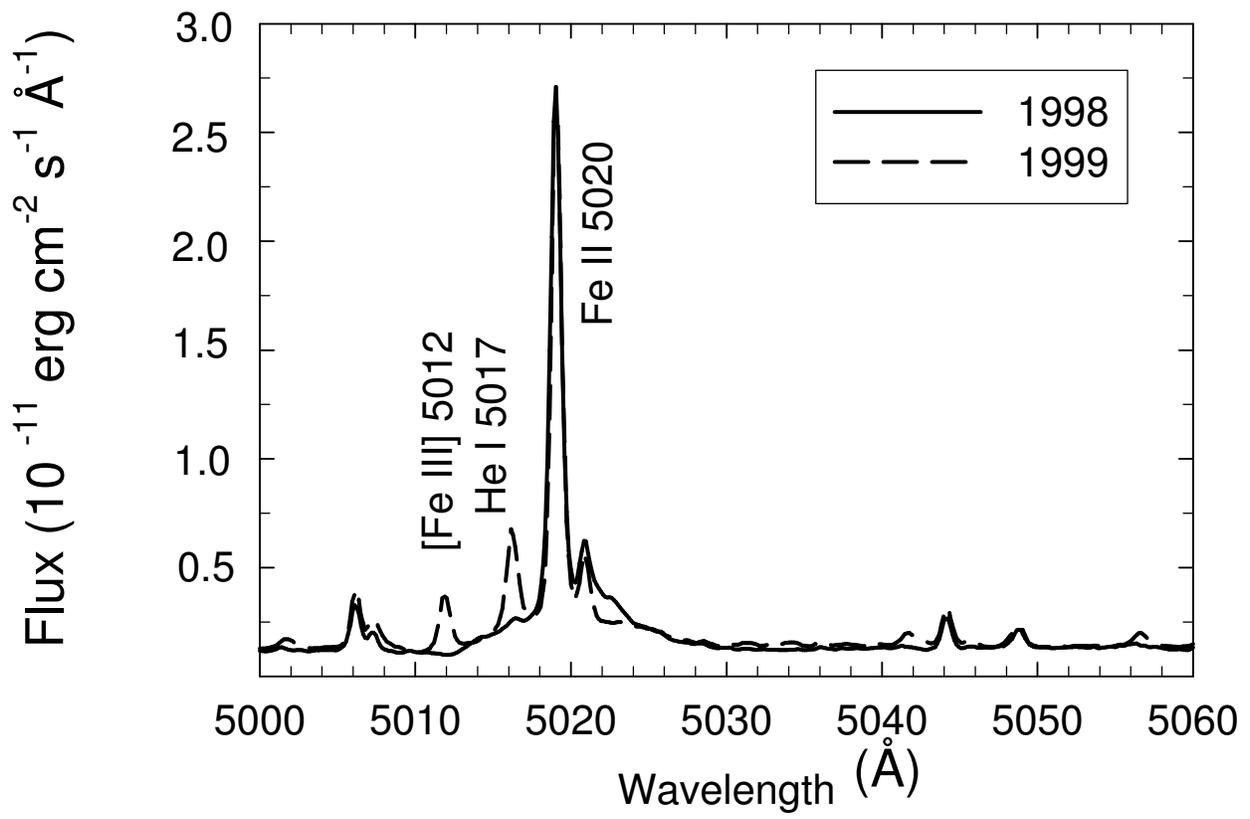

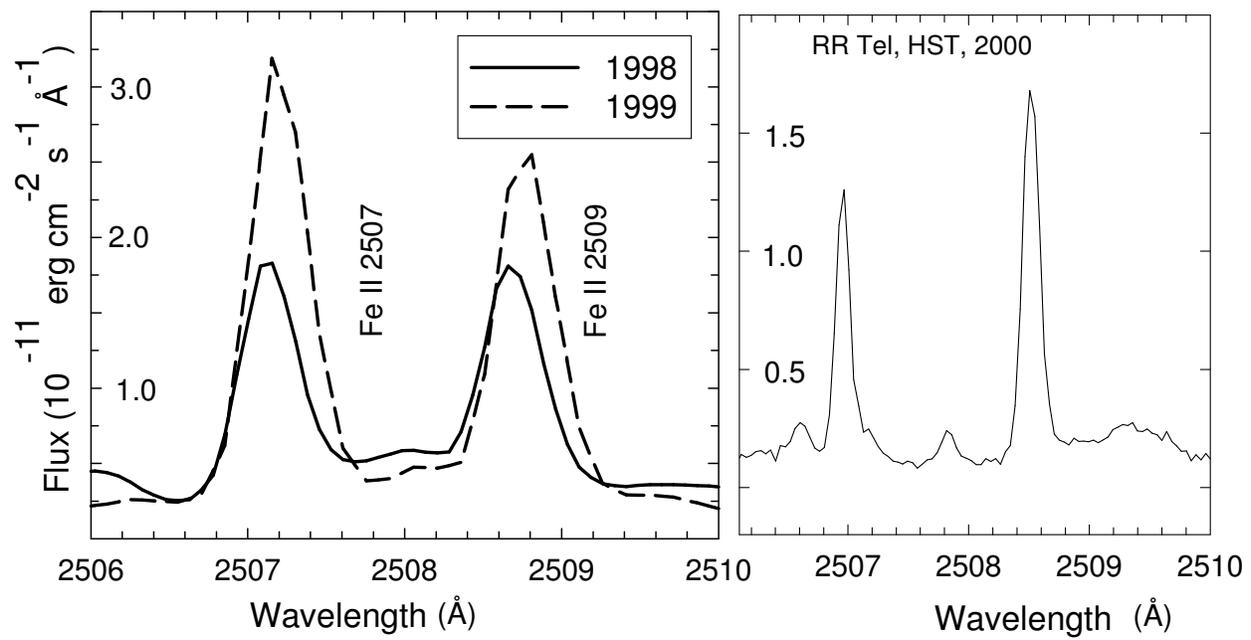

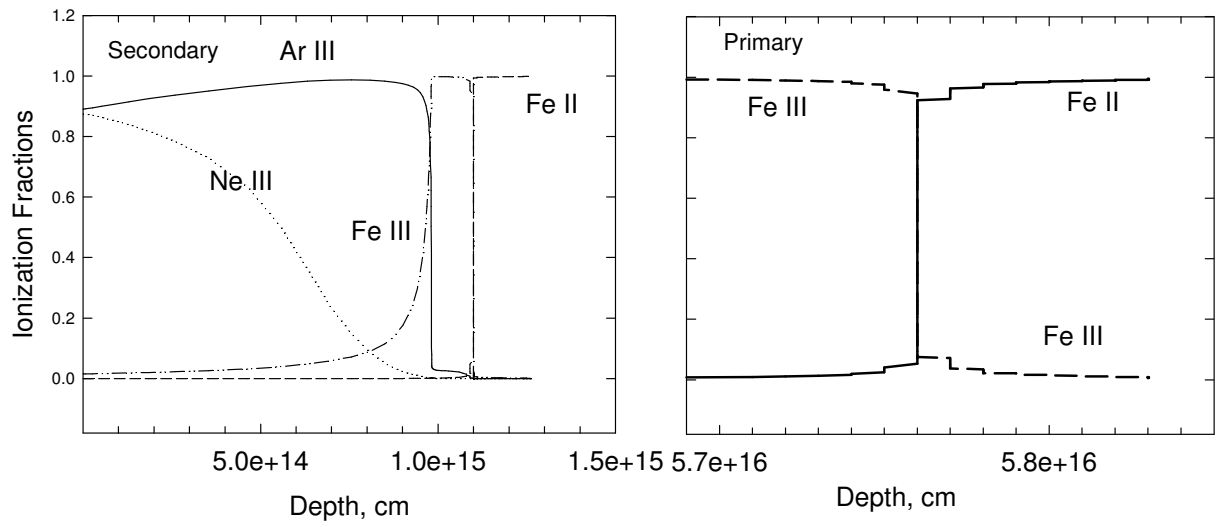